\documentclass{aa}

\usepackage{CJKutf8}
\usepackage{graphicx}
\usepackage[breaklinks,colorlinks,citecolor=blue,linkcolor=blue]{hyperref}
\usepackage{txfonts}
\usepackage{lscape}
\usepackage{epstopdf}
\usepackage{CJK}
\usepackage{float}
\usepackage[absolute]{textpos}
\usepackage{amsmath}
\usepackage{mathrsfs}

\usepackage{arydshln}

\usepackage{cases}

\newcommand{\xunchuancnname}{\begin{CJK*}{UTF8}{gbsn}刘训川\end{CJK*}}

\begin{document} 

   \title{Cloud2to3: Three-dimensional reconstruction via spherization}

   \author{Xunchuan Liu (\xunchuancnname)
          \inst{1}\thanks{liuxunchuan001@gmail.com} 
          \and Pak-Shing Li\inst{2}
          }

   \institute{Leiden Observatory, Leiden University, P.O. Box 9513, 2300RA
Leiden, The Netherlands
         \and
            Shanghai Astronomical Observatory, Chinese Academy of Sciences, 80 Nandan Road, Shanghai 200030, PR China
             }

   \date{Date xx xx}
 
\abstract{Inferring three-dimensional structures from two-dimensional maps remains a major challenge due to line of sight degeneracies. We present \texttt{Cloud2to3}, a flexible framework to reconstruct three-dimensional (3D) volumetric density fields from two-dimensional (2D) maps by generalizing the inverse Abel transform and the AVIATOR algorithm. The pipeline decomposes a 2D map into overlapping circles whose centers serve as the structural skeleton of each intensity slice. By linking these centers into a unified network, the 3D reconstruction problem is elegantly reduced to growing this skeletal tree along the line of sight. This work implements three benchmark strategies: a randomized approach, a deterministic approach, and a physically constrained model. While exact real space structural matching is degenerate due to a lack of local line of sight constraints, benchmarks show that both 2D and 3D density statistical properties are robustly preserved. The projected column density probability density function (PDF) remains invariant under multi-angle views, and testing against magnetohydrodynamic numerical simulation data confirms that \texttt{Cloud2to3} can recover the intrinsic volume density statistics. Overall, our framework preserves the continuity of primary morphological patterns, allowing features like filamentary junctions and dense core clustering to be qualitatively represented within the reconstructed volume.}

   \maketitle

\section{Introduction}
One of the main challenges in interpreting interstellar medium (ISM) observations is that, in most cases, only two-dimensional (2D) spatial information is available \citep[e.g.,][]{2006AJ....132.1158S, 2010A&A...518L.102A,2014A&A...571A..11P}. Although spectral observations offer additional constraints for interpretation \citep[e.g.,][]{2001ApJ...547..792D, 2019ApJS..240....9S,2024RAA....24b5009L,2026arXiv260618637L,2026ApJS..282...65Y}, a position-position-velocity (PPV) cube often fails to capture the critical information inherent in true three-dimensional position-position-position (PPP) data. For instance, the statistics on the volume density probability density function (PDF), particularly its variance and skewness, provide unique constraints on turbulence states, including the Mach number, forcing modes, intermittency, and the role of gravity in mass assembly \citep[e.g.,][]{1994ApJ...423..681V, 2010A&A...512A..81F,2010MNRAS.405L..56B,2017MNRAS.471.2730M, 2025arXiv251026077L, 2025arXiv250220458L, 2025NatAs...9.1366L, 2026arXiv260700872L}. Along with reconstructing this two-dimensional (3D) statistical information, recovering full 3D density fields enables the exploration of density fluctuations related to the initial conditions of star formation, evolution of fractal morphology, and detailed cloud excitation states through radiative transfer modeling \citep[e.g.,][]{1996ApJ...471..816E,2012MNRAS.423.2037H,2019A&A...622A..79J,2019A&A...622A..32L,2021ApJ...919...35Z,2022A&A...658A.140L,2025RAA....25b5020L,2026A&A...711A.220L}.

To address the loss of the third spatial dimension, traditional mathematical frameworks rely on restrictive geometrical assumptions. For distributions exhibiting spherical symmetry, the Abel transform \citep{Abel1826} provides a direct method to invert a 2D projected radial profile into its true 3D spatial counterpart. Because of this mathematical convenience, the Abel transform has been utilized extensively across various astrophysical fields to reconstruct idealized, symmetric structures \citep{1978ApOpt..17.3750G,1996RScI...67.2257B,2014A&A...562A.138R,2015ApJ...806..274L,2017A&A...604A..52B}. However, this idealized requirement for symmetry severely limits its utility when analyzing the actual ISM, where inherent anisotropy, filamentary fragmentation, and complex fractal geometry are widespread and often the primary focus of study \citep[e.g.,][]{2010A&A...518L.102A,2013A&A...555A.112P,2021MNRAS.508.2964A,2021ApJ...912..148L,2025RAA....25b5020L}. Due to the lack of more universal reconstruction methods, interpreting the 2D morphology of these complex, asymmetric ISM environments remains heavily reliant on human experience and qualitative intuition \citep[e.g.,][]{2012MNRAS.424.2442S,2021ApJ...912..148L,2021ApJS..257...51Y,2022ApJ...930L..22R,2023ApJ...948..109K,2023ApJ...943...90E,2025ApJ...985..230L}. Consequently, an efficient and adaptable method capable of inferring 3D structures from 2D images in a manner analogous to human visual intuition would be highly valuable for exploring the physics and morphology of the ISM.

An elegant advance in capturing structural intuition is the AVIATOR reconstruction algorithm \citep{2020A&A...633A.132H}, which successfully adapts the Abel transform to asymmetric geometries. Its layer by layer distance transform paradigm generalizes the classical concept of inflating a uniform circle into a sphere, allowing a 2D region with uniform surface density to be converted into an irregular 3D shape. While this approach relaxes the rigid constraint of global spherical symmetry, it introduces new structural restrictions by treating each discrete flux layer as a single object. Because the available degrees of freedom are strictly capped by the number of intensity contours, the method cannot reconstruct the complex volume density statistics essential for turbulence analysis. Furthermore, it prevents line of sight warping, forcing the recovered structures into rigid geometries that lack the flexible curvature observed in the projected 2D $x y$ plane. This structural rigidity challenges the modeling of complex hubs or filamentary webs where internal twisting is common, making AVIATOR difficult to apply as a universal solution for complex, multi scale interstellar fields. The underlying premise of this class of algorithms is to decompose a map into a structural layering of shapes of varying sizes, assigning a line of sight spatial scale to each localized component that mirrors its cross plane dimensions. This key concept of utilizing spatial scales as geometric proxies to enforce localized structural isotropy represents a fundamental foundation that has been widely adopted across other multi scale 3D density construction frameworks \citep[e.g.,][]{2014A&A...562A.138R,2025arXiv250917369L}.

To reconstruct three-dimensional (3D) structures or statistical properties from 2D maps, we introduce \texttt{Cloud2to3}. This framework generalizes the mathematical architecture of the Abel transform and the AVIATOR algorithm. Our method decomposes 2D flux slices into overlapping circles organized by a minimum spanning tree (MST). These circles inflate into spheres during 3D reconstruction. To determine $z$-axis coordinates, the MST is projected into 3D space under proper approaches. This fiducial implementation offers a flexible framework easily optimized by physical and kinematic constraints.

This work is organized as follows. In Sect.~\ref{sec_concept}, we introduce the basic notation and definitions. In Sect.~\ref{sec_decomposition}, we discuss the slicing and circular decomposition methods for segmenting the 2D map. In Sect.~\ref{sec_tree_expand}, we elaborate on different approaches for expanding the 2D structure into 3D space. In Sect.~\ref{sec_discussion}, we test our methodology on MHD numerical simulation data and discuss the corresponding limitations and future prospects of \texttt{Cloud2to3}. A brief summary is provided in Sect.~\ref{sec_summary}. The source code is publicly accessible online.\footnote{\url{https://gitee.com/liuxunchuan/cloud2to3/}}

\begin{figure}
    \centering
    \includegraphics[width=0.95\linewidth]{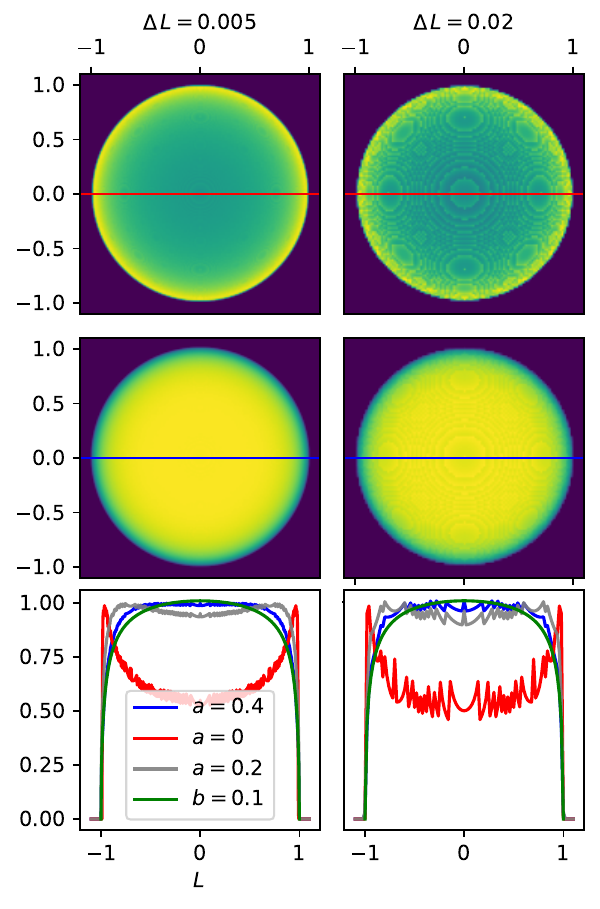}
\caption{Projected column densities under different structural configurations and grid resolutions. 
Upper panels: Reconstructed surface density $\Sigma_s$ for $a = 0$ (Sect.~\ref{sec_sphere_2to3}) using a finite grid cell size $\Delta L$. 
Middle panels: Same as the upper panels, but evaluated for a core parameter of $a = 0.4$. 
Lower panels: Radial profiles of $\Sigma_s(L)$ compared across different values of $a$. The green solid line represents the analytical model $\Sigma_r$ with $b=0.1$ (Eq.~\ref{eq_Sigma_suggested_sphererho}) under an infinitesimally small resolution limit ($\Delta L \to 0$). 
The left and right columns correspond to numerical resolutions of $\Delta L = 0.005$ and $\Delta L = 0.02$, respectively.}
\label{fig_sphere2to3}
\end{figure}

\begin{figure}[!t]
    \centering
    \includegraphics[width=0.95\linewidth]{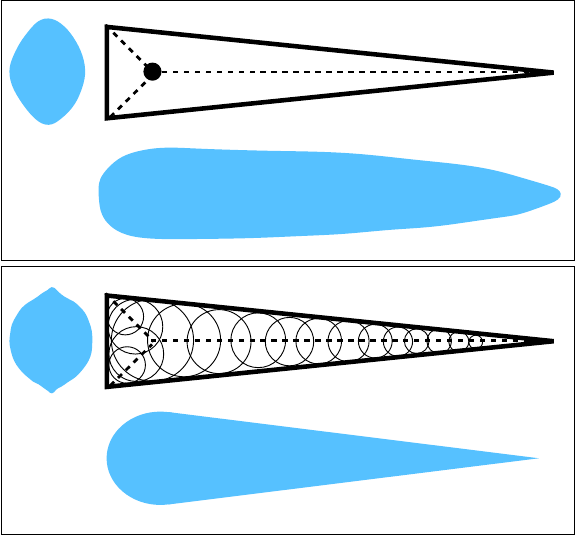}
\caption{Three-dimensional reconstruction for an idealized uniform triangle. 
Upper panel: Structure reconstructed using the AVIATOR algorithm \citep{2020A&A...633A.132H}. The dashed lines trace the interior angle bisectors, and the black dot marks the incenter, which corresponds to the global maximum of the distance transform (Sect.~\ref{sec_AVIATOR}). Shaded blue areas display the orthogonal projections of the 3D structure onto the $x$-$z$ (bottom) and $y$-$z$ (left) planes. 
Lower panel: Same as the upper panel, but reconstructed via the full spherization framework (Sect.~\ref{sec_concept}).}
\label{fig_triangle_circles}
\end{figure}

\section{Basic knowledge}\label{sec_concept}
In this work, the 2D projected radius is denoted as $L$, and the 3D physical radius as $r$. The 3D volume density and 2D surface (column) density are represented by $\rho$ and $\Sigma$, respectively. For uniformly gridded data, the 2D map elements are referred to as pixels, while the 3D grid cells are designated as voxels. The 2D map lies in the $x$-$y$ plane, with the $z$-axis serving as the default line of sight, oriented perpendicular to the $x$-$y$ plane. For a gridded volume, the indexing along the $x$, $y$, and $z$ directions is denoted by $i$, $j$, and $k$, respectively. 

A map is defined as a 2D column density (or integrated flux) map. A \textit{uniform shape} is defined as a simply connected 2D region characterized by a constant value inside its boundary and zero outside. A \textit{slice} (or \textit{layer}) is defined as a map containing one or more disconnected uniform shapes. Finally, an \textit{object} refers to a reconstructed 3D structure.

\subsection{Abel transform and individual spherization}\label{sec_sphere_2to3}
The inverse Abel transform \citep{Abel1826} dictates that a circularly symmetric 
2D projected distribution $\Sigma(L)$ uniquely maps to a spherically symmetric 
3D distribution $\rho(r)$ via the integral
\begin{equation}
\rho(r) = -\frac{1}{\pi} \int_r^\infty \frac{1}{\sqrt{L^2 - r^2}} \frac{d\Sigma(L)}{dL} \, dL.
\end{equation}
Consider a unit circle with radius $R = 1$ possessing a uniform column density 
distribution $\Sigma(L) = \Sigma_0 \cdot \Theta(1-L)$, where $\Theta$ is the 
Heaviside step function. The derivative of this profile with respect to the 
projected radius is given by a Dirac delta function
\begin{equation}
    \frac{d\Sigma(L)}{dL} = -\Sigma_0 \delta(L-1).
\end{equation}
Substituting this into the inversion formula yields the corresponding volume 
density distribution \citep[see also][]{2020A&A...633A.132H}
\begin{equation}
\rho_{s;\,\rm inf}(r) \propto \left( 1 - r^2 \right)^{-1/2} \quad \text{for } r < 1. \label{eq_origin+spherepho}
\end{equation}
This inversion can be forward-verified by integrating the volumetric profile 
along the line of sight ($z$-axis, where $r^2 = L^2 + z^2$)
\begin{align}
   \Sigma_{s;\,\rm inf}(L) &\propto \int_0^{\sqrt{1-L^2}} \left(1 - L^2 - z^2\right)^{-1/2} dz. \label{eq_Sigma_sphererho_int}
\end{align}
By applying the variable substitution $z = \sqrt{1-L^2} \cdot u$, Eq. 
\ref{eq_Sigma_sphererho_int} simplifies to
\begin{align}
   \Sigma_{s;\,\rm inf}(L) &\propto \int_0^1 \left(1 - u^2\right)^{-1/2} du = \arcsin(1) = \frac{\pi}{2}, \label{eq_Sigma_sphererho}
\end{align}
which recovers the constant, uniform column density distribution inside the unit circle.

\begin{figure*}[!t]
    \centering
    \includegraphics[width=0.95\linewidth]{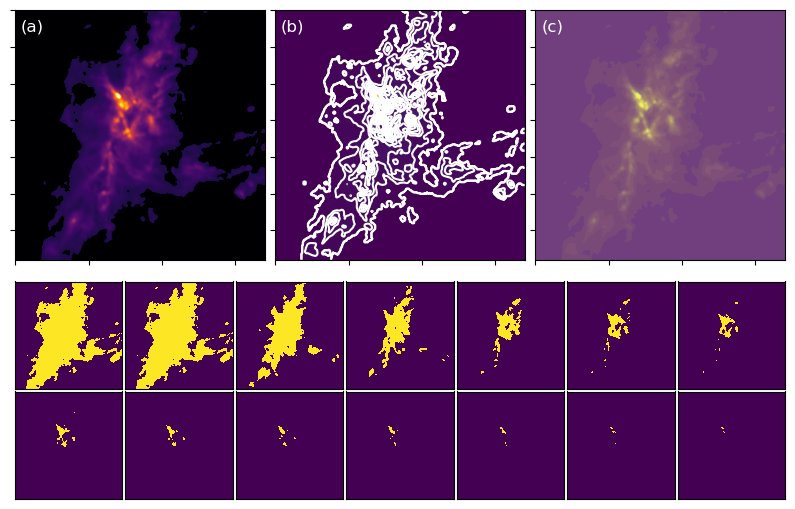}
\caption{Discretization of the example column density map (Sect.~\ref{sec_slicing}). 
(a) A $51\arcmin \times 51\arcmin$ cutout from the HGBS survey (Sect. \ref{sec_decomposition}). 
(b) Contour thresholds defining individual uniform slices. 
(c) Stacked projection of all 14 layers using an alpha transparency of 0.1 per slice. 
Lower panels: Spatial masks corresponding to each of the extracted slices.}
\label{fig_slice}
\end{figure*}

\subsection{Effect of finite resolution}\label{sec:finite_res}
Evaluating Eq.~\ref{eq_origin+spherepho} using voxel centers in a finite grid ($\Delta L$) produces an artificial column density $\Sigma_s$ with an over-brightened edge and depressed core (Fig.~\ref{fig_sphere2to3}). To circumvent separate boundary corrections \citep[e.g.,][]{2020A&A...633A.132H}, we propose a simplified, regularized 3D density profile
\begin{equation}
    \rho_s(r) \propto \left( 1 - r^2 + a\right)^{-1/2} \quad \text{for } r < 1,
    \label{eq_suggested_sphererho}
\end{equation}
where $a=0.4$ is an empirical parameter. 

In the continuum limit ($\Delta L \to 0$), line-of-sight integration yields
\begin{equation}
    \Sigma_{s}(L) \propto \arcsin\left( \sqrt{\frac{1-L^2}{1+b-L^2}} \right), \label{eq_Sigma_suggested_sphererho} 
\end{equation}
where $b \equiv a$, recovering Eq.~\ref{eq_Sigma_sphererho} when $a = 0$. Physically, Eq.~\ref{eq_suggested_sphererho} prevents boundary density divergence ($r \to 1$). This smoothly tapering profile better mirrors realistic interstellar clouds, where column densities are rarely strict top-hats and instead fall rapidly near the margins.

At finite resolutions, numerical integration of Eq.~\ref{eq_suggested_sphererho} yields a stable, flat inner $\Sigma_s$ distribution ($L < 0.7$) that is insensitive to voxel size (Fig.~\ref{fig_sphere2to3}). However, these discrete profiles are best fit by Eq.~\ref{eq_Sigma_suggested_sphererho} with an effective parameter of
\begin{equation}
    b \sim 0.1.
\end{equation}
This decoupling ($b < a$) occurs because discrete voxel summation inherently acts as a numerical smoothing filter, truncating the mathematical edge steepness. Provided the grid samples the bulk structure ($\Delta L \ll 1$), this numerical edge-smoothing remains stable, making $b \sim 0.1$ universally applicable across typical finite resolutions.
For structures with  $R\neq 1$, the spatial parameters $r$ and $L$ scale linearly as $r/R$ and $L/R$, respectively.

\subsection{The AVIATOR algorithm}\label{sec_AVIATOR}
The AVIATOR reconstruction framework \citep{2020A&A...633A.132H} decomposes a 2D intensity map into a series of hierarchical, discrete flux layers containing uniform shapes. Each shape is mapped onto a generalized coordinate system via a distance transform \citep[e.g.,][]{2021arXiv210603503S}. For any position $(x,y)$ within a uniform shape, the distance transform computes the Euclidean distance to the nearest outer boundary, denoted as $D_{xy}$. For geometric profiles like a triangle, $D_{xy}$ attains a unique local maximum at the intersection of the interior angle bisectors (the incenter; see Fig.~\ref{fig_triangle_circles}). Taking this local maximum as the geometric origin in the $x$-$y$ plane ($z=0$), the characteristic radius of the shape is defined as
\begin{equation}
    R = \max(D_{xy}),
\end{equation}
and the equivalent projected radius for any point within the shape is given by
\begin{equation}
    L_{xy} = R - D_{xy}.
\end{equation}
The generalized 3D physical radius $r$ for a voxel at coordinates $(x,y,z)$ is defined as \citep{2020A&A...633A.132H}
\begin{equation}
    r = \sqrt{z^2 + L_{xy}^2}.
\end{equation}
The spatial boundary of the reconstructed 3D object (upper panel of Fig.~\ref{fig_triangle_circles}) is bounded by the condition $r < R$, with its volumetric density assigned according to the  Abel inversion (Eq.~\ref{eq_origin+spherepho}).

The limitations of AVIATOR are inherent to this layer-by-layer distance transform paradigm. First, the reconstructed 3D volume of an individual flux layer is inherently oblate, with its line-of-sight ($z$-axis) extent strictly capped by its maximum trans-radial ($x$-$y$) dimension. Second, every reconstructed slice possesses strict mirror symmetry with respect to the midplane ($z = 0$). Third, while this mirror symmetry can be relaxed by translating the $z$-centers of distinct slices prior to final stacking, no self-consistent physical or mathematical constraint exists to determine these line-of-sight offsets. Fourth, because each discrete flux slice only undergoes rigid translation, the algorithm cannot capture differential $z$-axis warping or internal curvature within a single map level. This structural rigidity challenges the modeling of complex, multi-scale interstellar fields like twisted hubs or filamentary webs. Fifth, because the available degrees of freedom are strictly capped by the number of discrete intensity contours, the method cannot reconstruct the complex volume density statistics essential for turbulence analysis.

\begin{figure}
    \centering
    \includegraphics[width=0.95\linewidth]{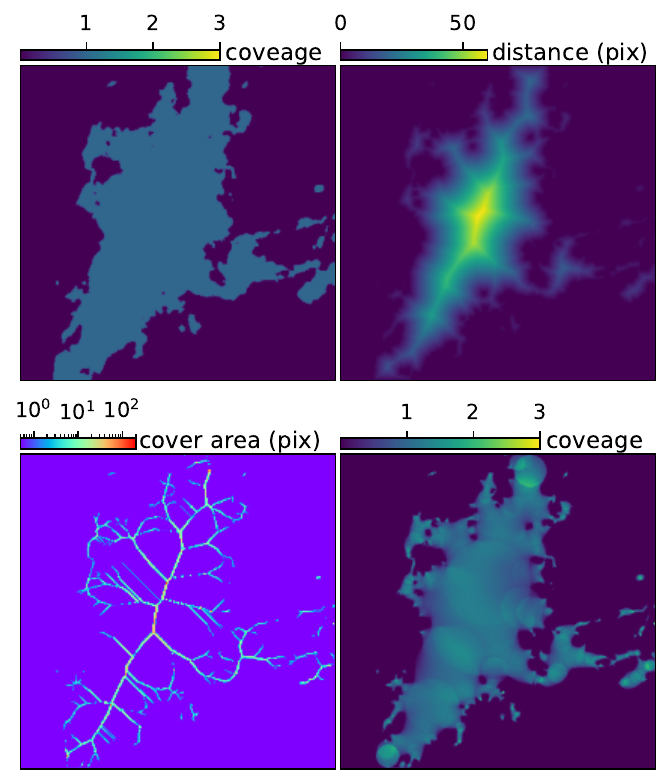}
    \caption{Upper left: The example slice of the example map enclosed by the slowest contour level shown in Fig. \ref{fig_slice}. Upper right: The distance transform of the example slice. Lower left: The medial axis of the example slice. The colors of the points on the medial axis mark their weights (Sect. \ref{sec_medialaxis}). Lower right: The slice reconstructed from the medial-axis points (Eq. \ref{eq_Omega1}).
    }
    \label{fig_onelayer_2dconstruct}
\end{figure}

\begin{figure*}
    \centering
    \includegraphics[width=0.95\linewidth]{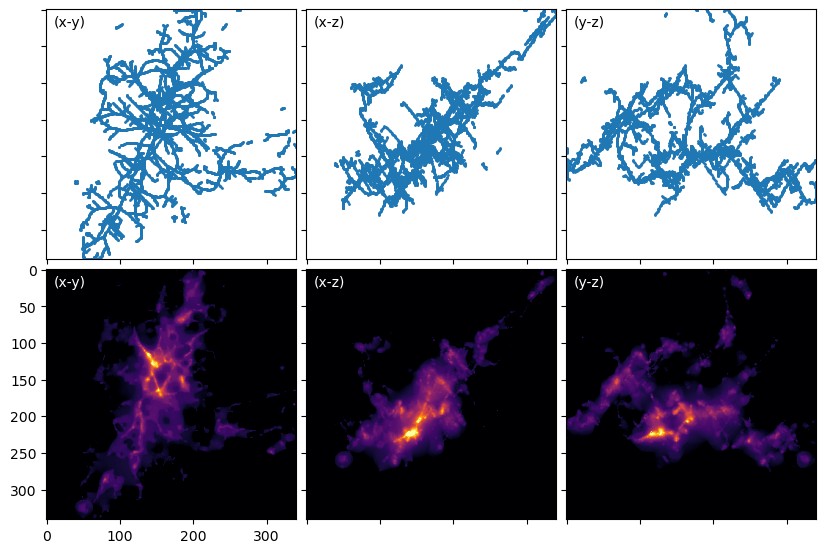}
    \caption{ The expanding tree of the medial-axis points and the constructed object, following the random approach described in Sect. \ref{sec_span3d_random}, projected onto three different major planes.
    }
    \label{fig_3D_reconstruct}
\end{figure*}

\section{Concept of full spherization}\label{sec_concept}
Filaments and fibers form the underlying backbone of molecular clouds within the interstellar medium (ISM) \citep[e.g.,][]{2010A&A...518L.102A, 2013A&A...555A.112P, 2025arXiv250210897L}. Reconstructing complex 3D clouds by extending a projected 2D tree-like skeleton into a 3D filamentary structure is therefore physically well-motivated. To achieve this, \texttt{Cloud2to3} first discretizes the intensity map into a series of uniform intensity slices. In contrast to traditional approaches, each slice here can accommodate multiple disconnected uniform shapes. More importantly, we extract a comprehensive, structured set of internal anchor points for each uniform shape rather than relying on a few discrete local maxima from a distance transform. These anchor points are extracted by implementing the medial axis as the structural skeleton of each shape,\footnote{This choice aligns with established filament-extraction techniques like FilFinder \citep{2015MNRAS.452.3435K}, reinforcing the medial axis as a robust backbone for structural representation.} which, when collected across different slices, comprise a massive node network defining the 2D skeletal tree of the map. Each anchor point is treated as a uniform circle characterized by an assigned mass weight and radius. The volumetric cloud is then constructed by inflating these individual nodes into spheres. Ultimately, the topological structure of this 2D tree serves as a geometric guide to map the nodes into a 3D tree, achieving an optimal balance between massive physical degrees of freedom and regularized, physically reasonable tree spanning.

The lower panel of Fig.~\ref{fig_triangle_circles} conceptually illustrates this spherization process for a uniform triangle, which is filled by overlapping circles centered along its angle bisectors (its medial axis). The loci of these centers trace the structural skeleton of the triangle. In its most basic implementation, each circle is inflated into a sphere via an Abel transform, with all spherical centers fixed to the midplane ($z=0$). This baseline configuration produces a 3D structure composed of three interconnected conical segments generated by rotating the triangle about its bisectors. Although this rigid setup shares the structural limitations of AVIATOR (Sect.~\ref{sec_AVIATOR}), the underlying medial-axis tree provides the mathematical flexibility to bend along the $z$-axis, offering crucial conceptual advantages for non-rigid reconstructions.

In essence, the \texttt{Cloud2to3} framework converts the challenge of 3D structural 
reconstruction into the mathematical spanning of a tree along the line of sight 
(detailed in Sects.~\ref{sec_decomposition} and \ref{sec_tree_expand}). The 
complex architectural patterns of the skeletal tree in the $x$-$y$ plane serve 
as a geometric template for its expansion along the $z$-axis. This line-of-sight 
spanning is highly flexible and can be further regularized by imposing external 
physical constraints, such as the assumption of hydrodynamic equilibrium or 
observational kinematic fields. Crucially, as we demonstrate in this work, even 
in the absence of additional physical constraints, the fiducial configuration 
of \texttt{Cloud2to3} yields a robust and stable recovery of 3D volumetric density 
statistics.

\section{Slicing and circular decomposition}\label{sec_decomposition}
This section describes our methodology for slicing 2D intensity maps (Sect.~\ref{sec_slicing}) and decomposing the uniform shapes into circles along their medial axes (Sect.~\ref{sec_medialaxis}). As a benchmark test for our methodology, we utilize a $51\arcmin \times 51\arcmin$ column density cutout of the Perseus molecular cloud from the Herschel Gould Belt Survey \citep[HGBS;][]{2010A&A...518L.102A, 2013A&A...550A..38P}, downsampled to $341 \times 341$ pixels (Fig.~\ref{fig_slice}a).

\subsection{Slicing}\label{sec_slicing}
An intensity map is first divided into discrete slices, each containing one or more uniform shapes. A direct approach is segmentation via ascending contour thresholds $I_l$, where $1 \leq l \leq l_{\rm max}$ and $I_1 < I_2 < \dots < I_{l_{\rm max}}$ \citep[e.g.,][]{2020A&A...633A.132H}. The spatial domain $\Omega_l$ of the $l$-th slice is defined by
\begin{equation}
    \Omega_l = \{ (x,y) \mid I_{xy} \ge I_l \},
\end{equation}  
where $\Omega_l$ can comprise multiple uniform shapes, such that $\Omega_l=\sum_m \Omega_{l,m}$. The corresponding slice height $h_l$ is given by
\begin{equation}
    h_{l} = \left\{ 
    \begin{aligned}
        &I_{l+1} - I_{l} & \text{for } l < l_{\rm max},\\
        & \frac{\int_{\Omega_{l_{\rm max}}} I_{xy} \,dx\,dy}{\int_{\Omega_{l_{\rm max}}}dx\,dy} - I_{l_{\rm max}} & \text{for } l = l_{\rm max}.
    \end{aligned}
    \right.
\end{equation}  
At low thresholds ($l \to 1$), emission regions are often diffuse and noisy; the map can be regularized via spatial smoothing prior to slicing \citep[e.g.,][]{2022ApJS..259...59L}.

Figure~\ref{fig_slice}b displays the contour boundaries of the example map evaluated at levels of $\exp\left(\ln(l)^{1.5}\right) \times 0.003 \, M_\sun\,\text{pixel}^{-1}$ for $1 \le l \le 14$. A linear accumulation of these 14 discrete layers (Fig.~\ref{fig_slice}c) faithfully reproduces the global morphology of the original map.

\begin{figure}[!t]
    \centering
    \includegraphics[width=0.95\linewidth]{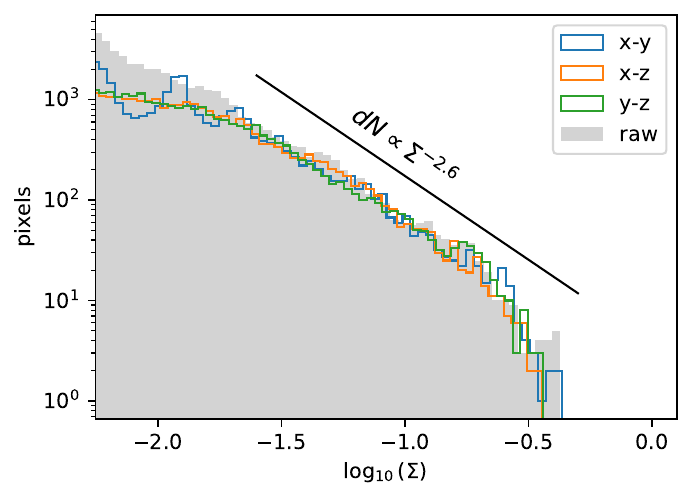}
    \includegraphics[width=0.95\linewidth]{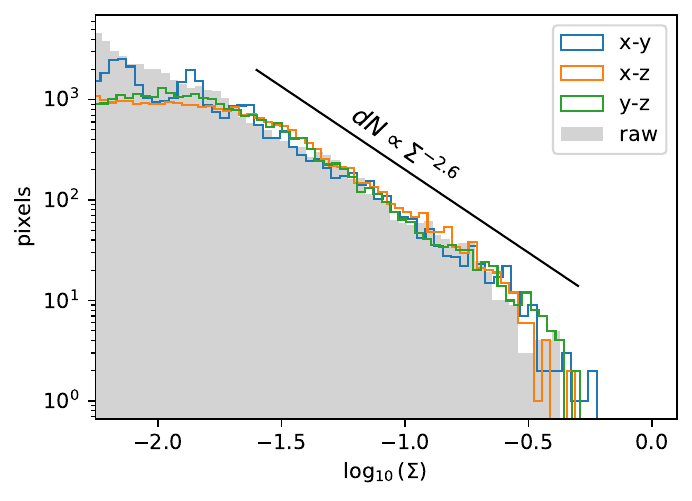}
    \includegraphics[width=0.95\linewidth]{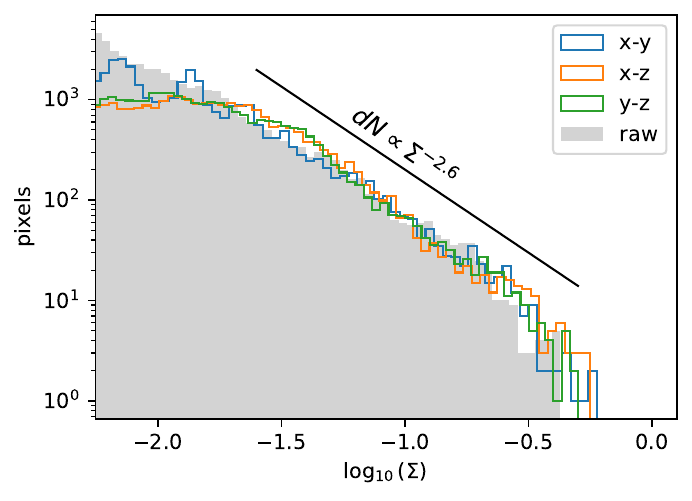}
    \caption{The gray area represents the $\Sigma$ PDF of the example map (Fig. \ref{fig_slice}). The stepped lines show the $\Sigma$ PDFs of the constructed object using the random approach (Sect. \ref{sec_span3d_random}), projected onto three different major planes (Fig. \ref{fig_3D_reconstruct}). The black line represents the power-law fit to the PDF at the high-$\Sigma$ end, with a power-law index of -2.6.
    Panels from top to bottom represent the
    construction of random approach (Fig. \ref{fig_3D_reconstruct}; Sect. \ref{sec_span3d_random}), deterministic approach (Fig. \ref{fig_fiducial_approach}; Appendix \ref{sec_fiducial}),
    and physical approcah (Fig. \ref{fig_physical_approach}; Appendix \ref{sec_ph_approach}),
    respectively.
    }
    \label{fig_randompdf}
\end{figure}

\subsection{Medial axis and weighting}\label{sec_medialaxis}
The decomposition process begins by computing the \textit{distance transform} of each shape, which assigns to each point $p$ within the shape its distance to the nearest boundary point. Let $\partial \Omega$ denote the boundary of the shape; the distance transform $D(p)$ is mathematically defined as
\begin{equation}
D(p) = \left\{
\begin{aligned}
    &\min_{q \in \partial \Omega} \| p - q \|, & \text{for } p \in \Omega, \\
    &0, & \text{for } p \notin \Omega,
\end{aligned} \right.
\end{equation}
where $\|\cdot\|$ represents the Euclidean norm. It quantitatively provides the proximity of each point to the boundary, which can be numerically computed using Python packages such as \texttt{scipy.ndimage.distance\_transform\_edt}.

Next, the \textit{medial axis} is extracted, corresponding to the set of points in the region that are equidistant from at least two boundary points with the same minimal distance. The medial axis $\mathcal{F}$ is mathematically defined as
\begin{equation}
  \mathcal{F} =  \left\{ p \mid \exists q_1, q_2 \in \partial \Omega, D(p) = \| p - q_1 \| = \| p - q_2 \|  \right\}.
\end{equation}
The medial-axis points form the skeleton of a uniform shape and can be numerically obtained via Python packages such as \texttt{skimage.morphology.medial\_axis}.

Finally, a \textit{watershed segmentation} technique is applied to determine the governing region of each medial-axis point. A point $q \in \Omega$ is governed by a marker $p_1 \in \mathcal{F}$ if, along the straight line connecting $p_1$ to $q$, the distance transform is non-decreasing, meaning for any $0 \leq t_1 \leq t_2 \leq 1$
\begin{equation}
   D\left(t_1 p_1 + (1 - t_1) q\right) \geq D\left(t_2 p_1 + (1 - t_2) q\right)>0. 
\end{equation}
If $q$ is influenced by multiple markers (e.g., $p_1$ and $p_2$), it is assigned to the marker with the smaller distance
\begin{equation}
\text{Governed}(q) = \begin{cases} 
p_1, & \text{if } D(p_1) \leq D(p_2), \\
p_2, & \text{if } D(p_2) < D(p_1).
\end{cases}
\end{equation}
Numerically, this procedure can be implemented using Python functions such as \texttt{skimage.segmentation.watershed}. The weight (or cover area) of a medial-axis pixel is defined as
\begin{equation}
    \mathcal{W}(p) = \text{count}\{q \mid q \in \Omega, \, \text{Governed}(q) = p\}.
\end{equation}
For each medial-axis pixel $p$, the key parameters are its coordinates $(i_p, j_p)$, the radius $R(p) = D(p)$, and the weight $\mathcal{W}(p)$.

The spatial domain of the first slice, $\Omega_1$ (the region enclosed by the lowest contour level of the example map), along with its distance transform and the corresponding medial-axis weight map, are shown in the first three panels of Fig.~\ref{fig_onelayer_2dconstruct}, respectively, to illustrate the decomposition procedure. The 2D structure of this single layer is forward-reconstructed in the $x$-$y$ plane by aggregating the medial-axis circles via
\begin{equation}
    \Omega_l^{\rm re}(i,j) = \sum_{p \in \mathcal{F}_l} \frac{\mathcal{W}(p)}{R^2(p)} \Sigma_s\left(\frac{\|(i,j) - (i_p, j_p)\|}{R(p)}\right), \label{eq_Omega1}
\end{equation}
where $\Sigma_s$ is evaluated using Eq.~\ref{eq_Sigma_suggested_sphererho} with $b = 0.1$, as shown in the lower right panel of Fig.~\ref{fig_onelayer_2dconstruct}. The reconstructed intensity field is not strictly uniform for a single slice. However, by summing the contributions from the medial-axis points across all hierarchical layers (Fig.~\ref{fig_slice}), the globally accumulated reconstruction,
\begin{equation}
    \Omega^{\rm re}(i,j) = \sum_{1 \le l \le l_{\rm max}} \Omega_l^{\rm re}(i,j),
\end{equation}
closely matches the original input map (as discussed in Sect.~\ref{sec_tree_expand}).

\section{Three-dimensional expanding and spherization}\label{sec_tree_expand}
By assigning a line-of-sight ($z$-axis) coordinate $k_p$ to each medial-axis point, the volume density field is reconstructed via
\begin{equation}
    \mathcal{O}(i,j,k) = \sum_l \sum_{p \in \mathcal{F}_l} \frac{\mathcal{W}(p)}{R^3(p)} \rho_s\left(\frac{\|(i,j,k) - (i_p, j_p, k_p)\|}{R(p)}\right), \label{eq_obj3d}
\end{equation}
where $\rho_s$ is evaluated using Eq.~\ref{eq_suggested_sphererho} with $a = 0.4$. Although the discrepancy between the multi-layered 2D forward-reconstruction $\Omega^{\rm re}(i,j)$ and the original observational intensity map is inherently small, we can apply a localized correction factor to eliminate residual differences, such that
\begin{equation}
   \mathcal{O}'(i,j,k) = \frac{I(i,j)}{\Omega^{\rm re}(i,j)} \mathcal{O}(i,j,k). \label{eq_obj3d_modif}
\end{equation}
Consequently, the foundational and most challenging task within this framework is establishing a physically and geometrically consistent method to assign the $z$-coordinates to the extracted medial-axis nodes.

\subsection{Spanning tree on the projected plane}\label{sec_spanningtree}
We assume that neighboring medial axis points ($\mathcal{F}$) on the projected plane tend to remain adjacent within the 3D volume. To characterize their connectivity, we construct a minimum spanning tree from all medial axis nodes based on their Euclidean distances in the $x y$ plane. Rather than building separate minimum spanning trees for each individual shape within each layer and cross matching them based on branch similarity, we adopt a simpler approach by constructing a single unified tree across all layers (also denoted as $\mathcal{F}$; see the upper left panel of Fig.~\ref{fig_3D_reconstruct}).

Within this tree structure, every node except the root ($p_0$) has a 
unique parent, and each non-leaf node has one or more children. The 
height of a subtree is defined as the length of the longest path from 
its root to a leaf. If multiple longest paths to a leaf exist within 
a subtree, one is selected as the \textit{best path}. A child node 
$p$ that lies on the best path of its parent node $q$ is designated 
as the \textit{best child} (where $p, q \in \mathcal{F}$), satisfying
\begin{equation}
    \operatorname{bestchild}(q) = p.
\end{equation}
Let $\mathcal{L}_p$ denote the length of the longest optimal path passing 
through a node $p$. A path is optimal if every node along it, except 
for the path's own starting node, is the best child of its parent. 
The coordinate increment of node $p$ relative to its parent $q$ is 
defined as
\begin{equation}
    \Delta p = (\Delta i_p, \Delta j_p, \Delta k_p) = (i_p - i_q, \, j_p - j_q, \, k_p - k_q), \label{eq_deltap}
\end{equation}
The root node of a longest optimal path $\mathcal{L}$ is denoted as 
$\operatorname{Root}(\mathcal{L})$. By definition, $\operatorname{Root}(\mathcal{L})$ 
cannot be a best child of any node.

\subsection{Random approach of $z$-coordinate expansion}\label{sec_span3d_random}
We further assume that the \(z\)-coordinate of the medial-axis points is distributed similarly to the \(x\) and \(y\) coordinates. This assumption motivates us to predict the \(z\)-coordinate of a node based on the \(x\) and \(y\) coordinates of that node and its parent node. The simplest form is
\begin{equation} \label{eq_dk_p}
    \Delta k_p = \cos(\theta_p) \Delta i_p + \sin(\theta_p) \Delta j_p.
\end{equation}
However, if \( \theta_p \) is a fixed value, e.g., \( \theta_p \equiv 0 \), it just yields
\begin{equation}
    k_p = i_p,
\end{equation}
leading to a constructed tree confined on the plane of $x=z$. 

In the random approach, we recommend the following step-wise assignment strategy for the orientation angles $\theta_p$
\begin{equation}
    \theta_p =
    \left\{
    \begin{aligned}
         & \theta_q + \frac{\pi}{\sqrt{\mathcal{L}_p}} \mathcal{N}(0, 1) && \quad \text{if } p = \text{bestchild}(q), \\
         & \theta_q + \frac{\pi}{2} \mathcal{U}(-1,1) && \quad \text{otherwise},
    \end{aligned}
    \right. \label{eq_random}
\end{equation}
where $q$ is the parent node of $p$, $\mathcal{N}(0, 1)$ represents the standard normal distribution, and $\mathcal{U}(-1,1)$ denotes a uniform distribution on the interval $[-1, 1]$. The physical motivation for this specific formulation is to enforce structural persistence along the main filamentary backbones, ensuring that the longest paths remain relatively straight, while allowing secondary short branches to behave as highly stochastic, jagged structures. The full 3D volume is subsequently generated via Eq.~\ref{eq_obj3d}. Orthogonal projections of this volumetric data onto the $x y$, $x z$, and $y z$ planes are then easily computed (see Fig.~\ref{fig_3D_reconstruct} for the three-view projections of the reconstructed skeletal tree and final 3D volume derived from the example map).


\subsubsection{Conservation of the column density PDF}
Due to the stochastic nature of the angle assignments in Eq.~\ref{eq_random}, individual 3D structural configurations remain highly flexible and inherently non-unique. However, this full spherization process preserves fundamental statistical parameters of the input map when viewing the constructed object from different angles. The probability density function of the column density (the $\Sigma$ PDF) is a cornerstone metric for ISM analysis, acting as an indispensable observational proxy for the 3D volume density PDF to track the physical evolutionary states of the compressive ISM affected by turbulence and self-gravity \citep[e.g.,][]{1994ApJ...423..681V, 2009ApJ...692...91G, 2015ApJ...811L..28B, 2025arXiv250210897L, 2025arXiv250220458L}. Both observational and numerical studies broadly attribute the emergence of high density power law tails in these distributions to gravitational collapse and mass assembly \citep[e.g.,][]{2019MNRAS.482.5233K, 2020A&A...635A..34K}.

Figure~\ref{fig_randompdf} displays the $\Sigma$ PDF of the example map compared directly against that of the re-projected map of the constructed object. The close alignment between these profiles indicates that the line of sight tree expansion and subsequent re-projection introduce negligible distortion to the global $\Sigma$ PDF, particularly across the high density regime. Linear power-law fits (in log scale) applied to the high-$\Sigma$ tails of both the original map and the reprojected maps yield identical power-law indices of $-2.6$, which well match the empirical valuations observed in active star-forming filaments \citep[e.g.,][]{2014MNRAS.445.2900S,2020A&A...635A..34K}.
This preservation occurs because the overlapping circles comprising a filamentary branch in 2D remain spatially coherent in 3D during the reconstruction process. Furthermore, this characteristic shape of the $\Sigma$ PDF remains well-preserved even when adopting alternative approaches for the 3D construction (Appendix \ref{sec_more_approaches}). These results strongly suggest that the $\Sigma$ PDF is a highly stable statistical feature of filament networks under geometric projection.



\section{Discussion}
\begin{figure}
    \centering
    \includegraphics[width=0.99\linewidth]{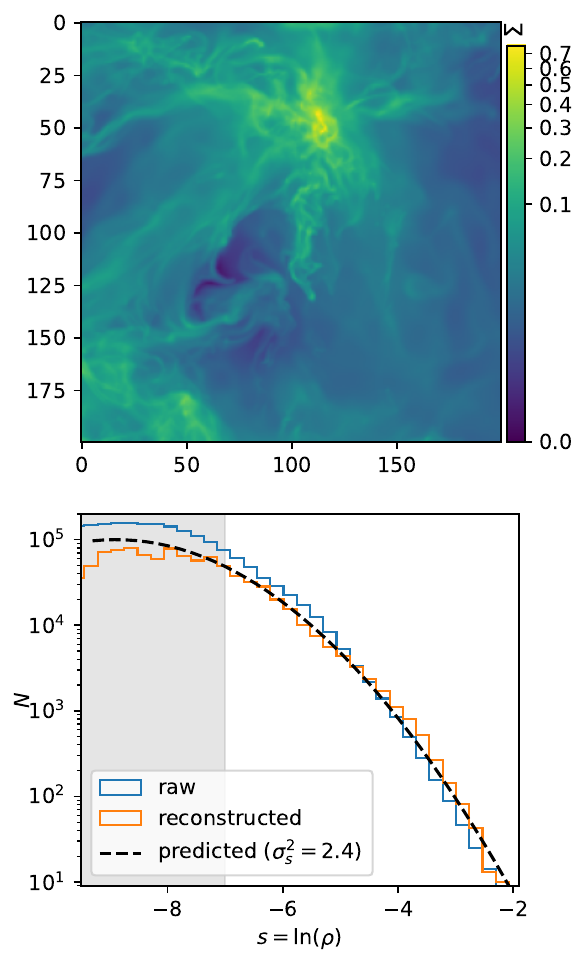}
    \caption{Reconstruction validation against MHD simulation data (Sect. \ref{sec_simu}). 
Upper panel: Projected surface density map of the subcube extracted from the turbulent cloud simulation catalog of \citet{2019MNRAS.485.4509L}, with intensities normalized to the global peak. 
Lower panel: Volume density PDFs of the raw simulation cube (black line) and the \texttt{Cloud2to3} reconstructed cube (blue line), compared against an analytical log-normal distribution profile evaluated at the corresponding Mach number. The lowest intensity contour threshold utilized for the surface slicing is 0.02. This boundary value corresponds to a baseline physical fidelity threshold of $0.02 / 200 = 0.001$ for the volumetric reconstruction, mapping directly to a $\ln(\rho)$ cutoff of $-7$ as indicated by the shaded grey region.}
\label{fig:simu}
\end{figure}
\label{sec_discussion}

\subsection{Test on numerical simulation data}\label{sec_simu}
To verify whether our full spherization framework can robustly recover true 3D volume density statistics, we apply the randomized construction approach to numerical simulation data. The upper panel of Fig.~\ref{fig:simu} shows the column (surface) density map, integrated along the $z$-axis, of a turbulence-dominated subcube with a resolution of $200^3$ voxels. This data is extracted from the synthetic magnetohydrodynamic (MHD) simulation of filamentary molecular clouds and active star formation presented by \citet{2019MNRAS.485.4509L}, which was conducted using the ORION2 adaptive mesh refinement code \citep{2012ApJ...745..139L}.
The 3D root-mean-square sonic Mach number ($\mathcal{M}$) of this selected subcube is 9.7. 

The volumetric density distribution of the 3D cube reconstructed from this projected surface density map closely resembles that of the raw, baseline simulation cube. Both distributions can be well-fitted by an analytical log-normal profile possessing a logarithmic density variance ($\sigma^2_s$) of 2.4 (lower panel of Fig.~\ref{fig:simu}). This value is in good agreement with the theoretical relation $\sigma_s^2 = \ln(1+b^2\mathcal{M}^2)$ evaluated with a turbulent driving parameter of $b = 1/3$ \citep{1994ApJ...423..681V, 2010A&A...512A..81F,2025arXiv250220458L}. This correspondence strongly indicates that the local cloud dynamics are either dominated by a purely solenoidal forcing mode mixture, or that the localized magnetic field architecture plays a major role in suppressing compressible density fluctuations. We conclude that \texttt{Cloud2to3} can robustly recover not only the 2D surface density statistical properties, but also the 3D volumetric statistical properties, providing a powerful avenue to systematically census the intrinsic variance of the 3D density distribution across observed molecular clouds.

\subsection{Limitations and future prospects}
The high degree of freedom inherent in the \texttt{Cloud2to3} framework allows for multiple valid 3D geometric interpretations from a single 2D input. A primary limitation is the handling of overlapping structures; filaments that are spatially distinct along the line of sight can blend in 2D projection, causing the medial axis tree to merge them into a single interconnected feature. To mitigate this ambiguity, future iterations could construct separate minimum spanning trees for independent hierarchical layers and subsequently cross-match branches based on topological similarity. 

The mathematical flexibility of the line-of-sight expansion can also be further regularized by incorporating complementary observational constraints. Integrating kinematic data from velocity cubes or multi-wavelength emissions into the node assignment rules will significantly improve the physical consistency of the reconstructed volumetric configurations \citep[e.g.,][]{2026MNRAS.550g1311X}. Furthermore, machine learning and deep learning techniques \citep[e.g.,][]{10.5555/2621980, 2023A&A...669A.120Z,2023ApJ...950..146X} offer promising pathways to automatically disentangle overlapping filamentary networks, optimize tree tracking, and guide the framework toward more realistic 3D configurations. Ultimately, applying this framework across large cloud samples will enable us to systematically chart the log-density ($\ln\rho$) variance of nearby molecular clouds, providing a vital tool for probing the competing influences of interstellar turbulence and self-gravity in star-forming environments \citep{2008ApJ...680..428G,2018Sci...360..635T,2025arXiv250220458L,2026arXiv260700872L}.


\section{Summary}
\label{sec_summary}
We have developed a flexible framework named \texttt{Cloud2to3} to reconstruct 3D volumetric density fields from 2D maps, generalizing the principles of the inverse Abel transform \citep{Abel1826} and the AVIATOR algorithm \citep{2020A&A...633A.132H}. The core pipeline decomposes a 2D map into layers of overlapping circles, connects their centers into a unified skeletal tree, and expands these nodes along the line of sight ($z$ axis). To perform this expansion, we provide three distinct operational strategies: a randomized approach, a deterministic approach, and a physically constrained model.

A key result of this work is that both the 2D projected and 3D volumetric density statistical properties are robustly preserved under the \texttt{Cloud2to3} reconstruction framework. Benchmarks show that the column density ($\Sigma$) PDF remains invariant across different coordinate expansion strategies and multi angle projections. Furthermore, validation against synthetic magnetohydrodynamic simulation data confirms that the framework successfully recovers the intrinsic 3D log normal density variance. While uniquely matching the exact real space 3D structure is mathematically degenerate, \texttt{Cloud2to3} preserves the continuity of primary morphological patterns, allowing features such as filamentary junctions, dense core clustering, and internal twisting to be qualitatively represented in the reconstructed volume. This framework offers an extensible platform where incorporating supplementary kinematic or multi wavelength observations can further regularize the tree spanning process.

\begin{acknowledgement}
X. Liu acknowledges the support of the Strategic Priority Research Program of the Chinese Academy of Sciences  under Grant No. XDB0800303.
We thank Dr. Xiaofeng Mai for helpful discussions during the early stages of this work.
\end{acknowledgement}

\bibliographystyle{aa}
\bibliography{Cloud2to3}

\begin{appendix}
\section{Alternative construction approaches}\label{sec_more_approaches}
\subsection{Deterministic approach}
\label{sec_fiducial}
The randomized approach described in Sect.~\ref{sec_span3d_random} relies on stochastic variables, which introduces structural non uniqueness. To establish a predictable mapping, we introduce a deterministic alternative. While this geometric approach does not model explicit cloud dynamics, it offers a consistent, reproducible, and explainable mathematical transformation to map a 2D intensity profile into a 3D volume.

The core logic of this approach is to use the local twisting or bending of a filament branch in the 2D sky plane to systematically guide its curvature along the line of sight. For each longest best path $\mathcal{L}$, we first characterize its global, average orientation in the $x y$ plane by an angle $\theta_{\mathcal{L}}$, which corresponds to a directional vector given by $(\cos \theta_{\mathcal{L}}, \sin \theta_{\mathcal{L}})$. To determine this main baseline direction for complex, realistic shapes, we implement a principal component analysis algorithm \citep[PCA; e.g.,][]{2017A&A...599A.100G} via
\begin{equation}
    \theta_{\mathcal{L}} = \operatorname{atan2}( v_y,\ v_x ),
\end{equation}
where $(v_x, v_y)$ represents the components of the major eigenvector calculated from the coordinates of the nodes comprising $\mathcal{L}$. Next, the local position angle of any individual node $p$ relative to its path origin node (${\rm Root}(\mathcal{L}_p)$) of the optimal path it belongs to ($\mathcal{L}_p$) is denoted as
\begin{equation}
    \theta_{p;\mathcal{L}_p} = \angle \left( p, {\rm Root}(\mathcal{L}_p) \right).
\end{equation}
The angular deviation of this localized position angle from the global path orientation is given by
\begin{equation}
    \Delta \theta_{p, \mathcal{L}_p} = \theta_{p;\mathcal{L}_p} - \theta_{\mathcal{L}_p},
\end{equation}
which quantifies the structural tortuosity, or physical twisting, of $\mathcal{L}_p$ in the plane of the sky. We then translate this 2D twisting directly into line of sight structural variations by calculating a projection angle $\theta_p$ for each node via
\begin{equation}
    \theta_p = \theta_{\mathcal{L}_p} - f \Delta \theta_{p, \mathcal{L}_p} + \frac{\pi}{2}, \label{eq_fiducial_tp}
\end{equation}
where $f$ is a scaling parameter that governs the deterministic expansion. The value of $f$ remains constant throughout a single reconstruction, with a default value of $f = 2$ adopted in this work. Using $\theta_p$, the $z$ axis coordinate increment $\Delta k_p$ between node $p$ and its parent is calculated through Eq.~\ref{eq_dk_p}. Under this formulation, the algorithm translates sharp 2D directional changes in the map plane into proportionally larger line of sight coordinate displacements in 3D space, mapping localized bends into three dimensional curves. 

\begin{figure*}
    \centering
    \includegraphics[width=0.8\linewidth]{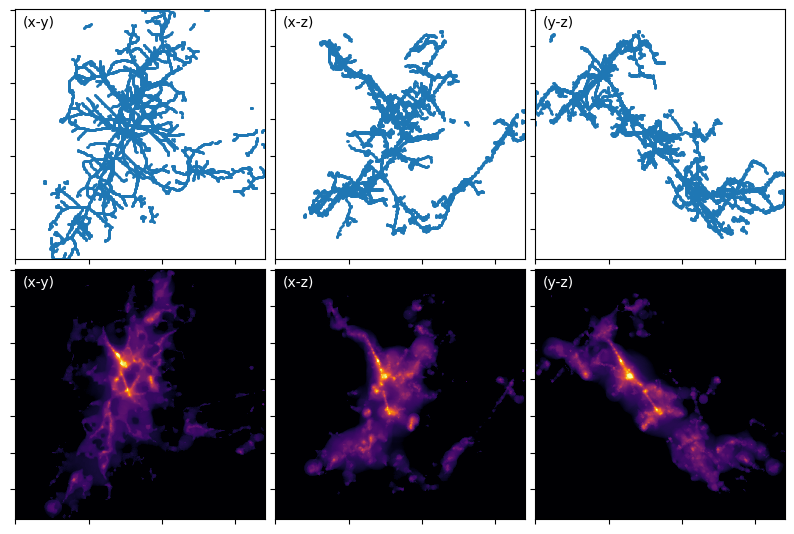}
    \caption{The expanding tree of the medial-axis points and the constructed object, following the deterministic approach described in Appendix \ref{sec_fiducial}, projected onto three different major planes.\label{fig_fiducial_approach}}
\end{figure*}

Figure~\ref{fig_fiducial_approach} presents the 3D volume reconstructed from the example map using this deterministic approach with $f = 2$. Visually, the spatial continuity of primary morphological patterns, such as filamentary junctions, dense core clustering, and web like twisting, is qualitatively preserved when viewing the constructed volume from alternative projection angles. Crucially, this geometric consistency ensures that the projected column density ($\Sigma$) PDF remains nearly invariant under multi angle projections (Fig.~\ref{fig_randompdf}), confirming the statistical stability of the \texttt{Cloud2to3} deterministic pipeline.

\begin{figure*}[!h]
    \centering
    \includegraphics[width=0.8\linewidth]{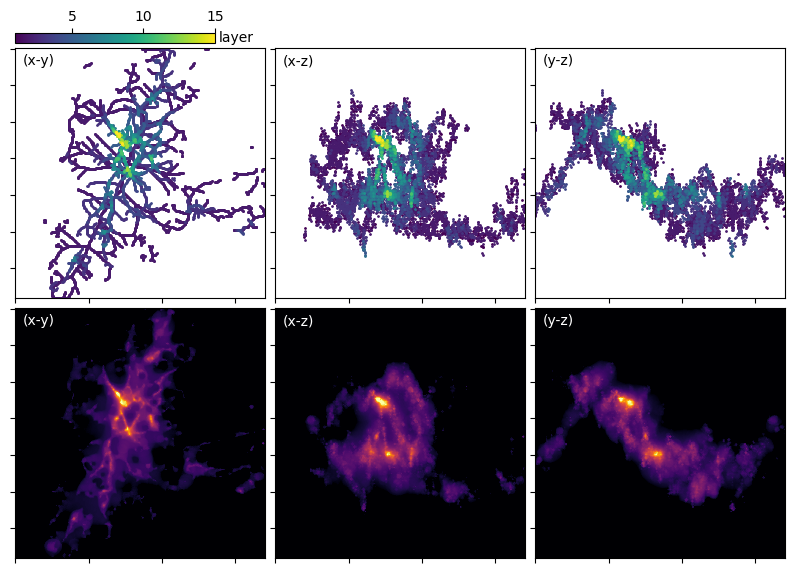}
    \caption{The expanding tree of the medial-axis points and the constructed object, following the physical approach described in Appendix~\ref{sec_ph_approach}, projected onto three different major planes. The color in the panels of the first row represents the layers to which the nodes belong. \label{fig_physical_approach}}
\end{figure*}

\subsection{Physical Approach}
\label{sec_ph_approach}
The randomized approach (Sect.~\ref{sec_span3d_random}) and the deterministic approach (Appendix~\ref{sec_fiducial}) impose minimal physical boundaries beyond mapping the skeletal tree topology to filamentary networks. Consequently, they operate primarily as pure image-based transformations. To anchor the reconstruction in realistic cloud physics, we must introduce reasonable dynamical regularizations during the line-of-sight tree expansion. For example, if a PPV cube is available, the observed centroid velocities of the medial axis points could map directly to $z$-coordinates, or the local line widths could scale the coordinate increments $\Delta k_p$ instead of fixing them to unity.

In this work, we focus on a fundamental physical constraint by requiring the reconstructed 3D volume to establish a state of quasi-hydrodynamic equilibrium. For simplicity, we assume the cloud possesses a uniform temperature $T$. Under this isothermal assumption, the probability of a given gas component settling at a specific line-of-sight coordinate $z$ obeys a Boltzmann distribution given by
\citep[e.g.,][]{1978ppim.book.....S}
\begin{equation} 
P^{\rm ph}(z;p) \propto \exp\left( \frac{-\phi_p(z)}{T} \right),  
\label{eq_physics_constraint}
\end{equation}
where $\phi_p(z)$ is the local gravitational potential at node $p$, and $T$ serves as a shorthand notation for the thermal energy scale $kT$ (with $k$ denoting the Boltzmann constant).

To compute this potential numerically, we first consider an idealized unit sphere with unit mass and unit radius whose interior density distribution follows Eq.~\ref{eq_origin+spherepho}. The mass enclosed within an internal radius $r < 1$ is analytically expressed as
\begin{equation}
    M(r) = \frac{2}{\pi} \left( \arcsin r - r \sqrt{1 - r^2} \right).
\end{equation}
By omitting the gravitational constant, the corresponding gravitational potential profile of this unit sphere is given by
\begin{equation}
\phi(r) = \left\{
\begin{aligned}
    &-\frac{2}{\pi} \left( \sqrt{1 - r^2} + \frac{\arcsin r}{r} \right) & \text{for } r < 1, \\
    &-\frac{1}{r} & \text{for } r \ge 1.
\end{aligned}
\right.
\end{equation}
For two overlapping spheres with arbitrary radii $R_1$ and $R_2$ and unit masses, we implement a structurally softened potential to approximate their mutual gravitational potential energy while neglecting inner self-gravity:
\begin{equation}
    \phi(d;R_1,R_2) = -\frac{1}{\sqrt{d^2 + 0.4R_1^2 + 0.4R_2^2}},
    \label{eq_phi_sp_approx}
\end{equation}
where $d$ is the spatial distance separating the two sphere centers. This softening formulation is critical because it mathematically prevents unphysical potential singularities as $d \to 0$. The total gravitational potential $\phi_p$ at any anchor node $p$ is then securely evaluated by summing the contributions from all other nodes across the tree structure via
\begin{equation}
    \phi_p = \sum_{q \neq p,\, q \in \mathcal{F}} \phi_{p;q},
\end{equation}
where the individual pair potentials $\phi_{p;q}$ are computed using Eq.~\ref{eq_phi_sp_approx}.

Concurrently, the geometric constraint imposed by the underlying medial axis tree structure is defined as  
\begin{equation}
    P^{\rm tr}(z;p) \propto \exp\left( -\frac{(z-k_q)^2}{2 ( \Delta i_p^2 + \Delta j_p^2)} \right), \label{eq_tree_constraint}
\end{equation}  
where $q$ denotes the unique parent node of $p$, and the spatial parameters $k_q$, $\Delta i_p$, and $\Delta j_p$ match the coordinate definitions established in Eq.~\ref{eq_deltap}. To seamlessly merge the regularizing cloud physics with the structural skeletal boundaries, we multiply these independent probabilities to construct a unified, combined probability distribution function:
\begin{equation}
  P^{\rm cb}(z;p) \propto P^{\rm ph}(z;p) P^{\rm tr}(z;p). \label{eq_physical_combine}
\end{equation}
This combined framework acts as an adaptive filter across different spatial scales. For prominent nodes with large radii, the gravitational potential term in Eq.~\ref{eq_physics_constraint} permits a broad distribution of line-of-sight steps $\Delta k_p$, allowing large-scale structures to expand freely along the $z$-axis. Conversely, for narrow filamentary paths composed of small-radius nodes, the tree constraint in Eq.~\ref{eq_tree_constraint} would natively trigger a chaotic random walk along the line of sight; however, the gravitational term in Eq.~\ref{eq_physics_constraint} suppresses this stochasticity, channeling the nodes into smoothly curved 3D filamentary vectors.

In practice, we apply this physical approach to the example map by implementing a simulated annealing protocol, where the operational temperature $T$ is gradually lowered from $30\text{~K}$ down to a typical molecular cloud baseline of $15\text{~K}$ over successive reconstruction iterations. The resulting 3D configuration is displayed in Fig.~\ref{fig_physical_approach}. As expected from the underlying governing equations, the spatial scatter along the $z$-axis is noticeably larger within the extended, lower-density layers than within the highly compact, peak-density layers. Crucially, the coherent continuity of the filamentary backbones is preserved. Furthermore, the projected column density ($\Sigma$) probability density function (PDF) remains invariant under multi-angle projections (Fig.~\ref{fig_randompdf}), demonstrating that incorporating explicit gravitational equilibrium criteria preserves the fundamental statistical robustness of the \texttt{Cloud2to3} pipeline.
\end{appendix}

\end{document}